\documentclass[12pt,a4paper]{article}
\usepackage{graphicx,color,epsfig}

\begin{document}
\large
\section*{
\centerline{Traffic flow stochastic model $(2\times 2)$}
\centerline{with discrete set of states}
\centerline{and continuous time}
}
\vskip 1cm
{\it Buslaev A.P., Tatashev A.G., Yashina M.V.}
\vskip 1cm

\centerline{\parbox{11cm}{\bf The present paper
proposes a stochastic model of the traffic flow.
This model has a discrete set of states and the continuous time.
The model is a generalization of the discrete stochastis model that
has been considered in a previous paper of this authors collective, for the
case of a mixed traffic flow which consists of "fast" and "slow" vehicles.
Moreover the continuous time model allows to developed the manner for
estimation of characteristics of the traffic flow.}}
\vskip 1cm

\subsection*{1. Introduction}

The analysis of the traffic flow and problems of the traffic control
becomes more pressing last years. Deterministic (hydrodynamic) models, that
do not take into account some substantial factors as availability
of several lanes, the different characteristics of traffic participants
and so on, parallel with stochastic models are considered. In stochastic
models the road is treated as a set of cells that proportional to the
dynamic distance and a discrete time unit. The traffic flow velocity
consists of the deterministic component connected with dynamic distance and
a stochastic component that individual behavior into account. So it has been 
studied the particles of different types
movement on a set of cells. The dimension of this set is $N\times m,$ $N>>1,$
$m>1.$

The discrete model of traffic flow on a one-lane road has been considered in
[1,2]. The road is represented by a set of subsequent cells. In each cell
there is one or no particle (vehicle). The transitions of particles are
possible only at discrete time units $n\Delta.$
Suppose that a cell is occupied by
a particle and the cell following ahead is vacant. Then with probability
depending on the particle type (vehicle velocity) transition of
the particle forwards to the following cell occurs. The steady
characteristics of the considered system have been investigated. The
multilane simulation models have been considered in [3]. The
analysis of traffic flow characteristics on the basis of simulation has been
presented in [4, 5]. Traffic flow mathematical models have been investigated
in [6, 7]. A deterministic model has been investigated in [6]. It has
been considered a stochastic model in terms of the queuing theory in [7].

A two-lane traffic flow discrete stochastic model, that is a generalisation
of the models, considered in [1,2], has been developed in [8].
In this model a road is represented by two closed sequences of cells
(lanes).
The transitions of particles are possible only at
discrete time units $n\Delta.$
If the cell following ahead is busy and the two
adjancent cells are empty
then with the appropriate probability the particle is rebuilt in adjacent
lane with propulsion. In
[8] the approximate method has been elaborated for evaluation of the steady
state probabilities of considered system fragments and such characteristics
as the particle flow rate (traffic flow rate) and the changes of lane rate.
The exactness of proposed approximations has been estimated by means of
simulation modelling.

{\it  In the present paper the model is investigated which differs from
the model considered in [8] by that in this model transitions of particles
occur at arbitrary instants with an assigned intensity. There are two types
of particles characterized by the intensity of particle transitions.} Thus
the model takes into account the availability in the traffic flow of "fast"
and "slow" vehicles.

The continuous time model is the limit case of the discrete time model
when  $\Delta \to 0$ and accordingly the transition probabilities $p\to 0.$

The method used for calculation of the studied characteristics is
similar to the method proposed in [8]. It is proved that the continuous
time model allows to deal with less complicated calculations than the
appropriate discrete time model. It is  explained by that {\it in a
continuous time model every change of  states of a cells set is realized
by a transition of the only particle.}

\subsection*{2. Model description}

Suppose that there are $N\times 2$ cells. Two indexes correspond to each
cell. The value of the first index can be equal to
$1,2,\dots,N$ and the second index (lane index) can be equal 1 or 2.
So the pair of numbers $(i,j),$ $i=1,2,\dots,N,$ $j=1,2,$ corresponds to
the cell. There are $M_1$ particles of the first type and $M_2$ particles
of the second type. At each instant each particle occupies one
cell and no cell contains more than one particle.

Suppose that the value of the first index, which equal to $N+1,$ is
identified with 1 and the second index, which is equal to $j+1,$ is
identified with 1 when $j=2.$

The particle movement is submitted to the next rules. {\it If at
time $t$ a particle of the $k$th type $(k=1,2)$ occupies the cell $(i,j)$
and the cell $(i+1,j)$ is vacant then during the time interval
$(t,t+\Delta)$ with probability $\mu_ k\Delta+o(\Delta), \Delta \to 0,$
the particle passes from the cell $(i,j)$ to the cell $(i+1,j)$ (i.e.
the particle move forward remaining in the same lane). If the particle of
the $k$th type occupies the cell $(i,j)$, the cell $(i+1,j)$ is busy, and
the cells $(i,j+1),$ $(i+1,j+1)$ are empty then during the time interval
$(t,t+\Delta)$ with probability $\mu_k \Delta+o(\Delta)$ the particle moves
to the cell $(i+1,j+1)$, i.e. moves forward with change of lane. In the
other cases the particle is not replaced. The probability of two or more
transitions during the time interval is the infinitesimal of higher oder
than $\Delta.$}

\subsection*{3. Evaluation of the steady state probabilities}

Let us consider the four cells with the first index, which is equal to 1 and
2. For approximate calculation of the steady states of the four cells we use
method similar to the method which has been elaborated in [8].

{\it The mean idea of the approximated method is that the considered
four cells are described by a Markov process with discrete set of state and
continuous time. The transition intensity for each pair of the process
is calculated provided that in every cell, adjancent to the four cells,
$((N,1),$ $(N,2),$ $(3,1),$ $(3,2))$ with probability  $\rho_1=m_1/(2N),$
that is independent on the other cells states, there is a particle of the
first type, with the probability $\rho_2=m_2/(2N)$ the cell is occupied
by a particle of the second type and with probability $1-\rho$
$(\rho=\rho_1+\rho_2)$ the cell is empty.}


Taking into account that the configurations of particles, symmetrical
respecting to line between the lanes, are characterized by just the same
probabilities, we compound them to one state. There are 45 different states
of the four cells shown in fig. 1. Suppose that equality
$\theta(i,j)=k,$ $k=1,2,$ means that the cell $(i,j)$ contains the
particle of the $k$th type. The equality $\theta(i,j)=0$ means that the
particle $(i,j)$ is empty. Then, for example, the states $E_1$ and $E_2$
are described as $E_1=\{\theta(1,1)=\theta(1,2)=\theta(2,1)=\theta(2,2)=0\},$
$E_2=\{\theta(1,1)=\theta(1,2)=\theta(2,1)=0,\
\theta(2,2)=1\}\bigcup\{\theta(1,1)=\theta(1,2)=\theta(2,2)=0,\
\theta(2,1)=1\}.$

\begin{figure}[h]
\centerline{\includegraphics[width=14cm]{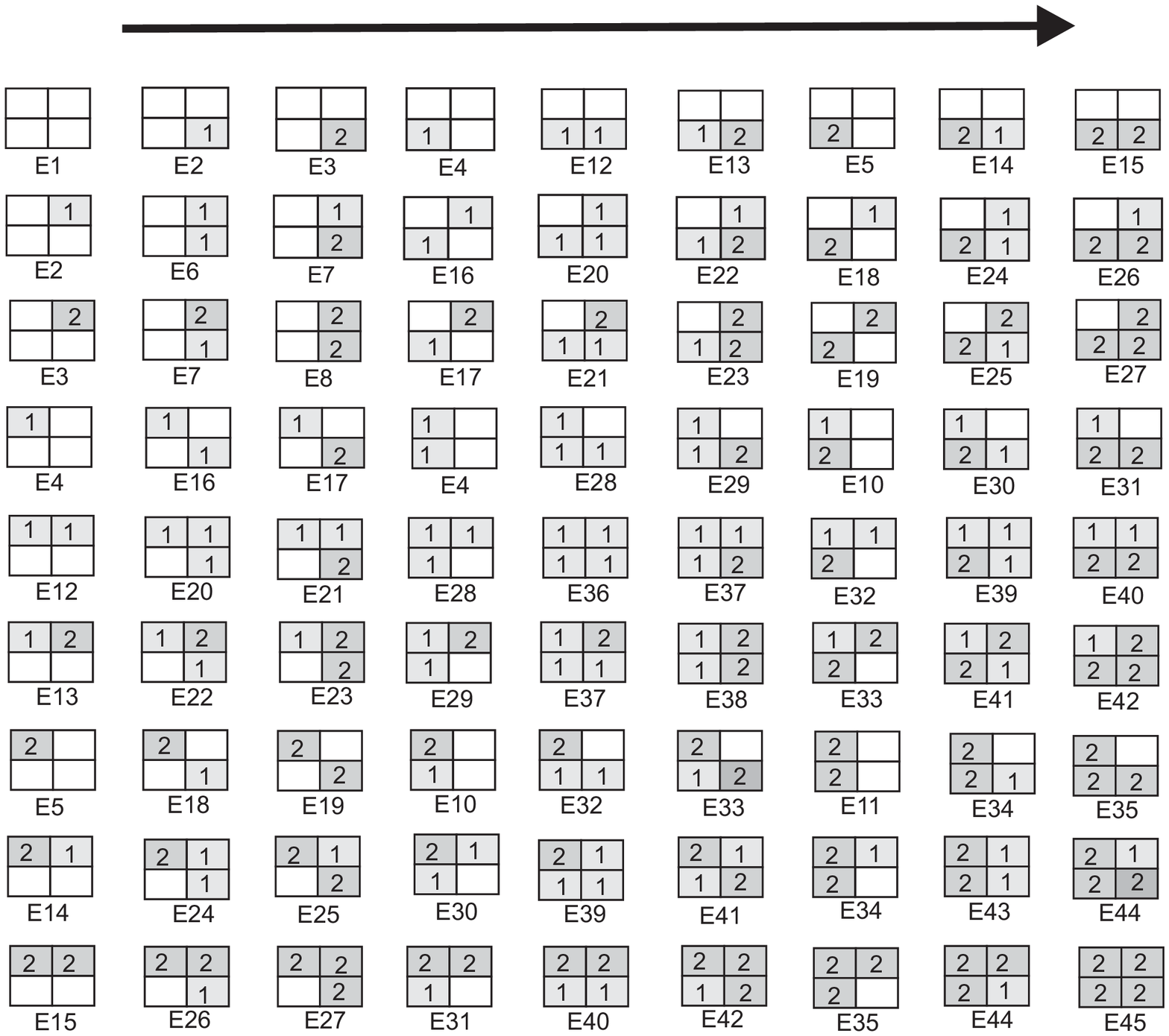}}
\caption{States of the four cells}
\label{sost4}
\end{figure}

Let $P_i(t)$ be the probability that at time the state of the four cells is
$E_i.$ The considered Markov process satisfies the conditions of the
ergodicity [9]. So the steady state probabilities $p_i$ exist,
$p_i=\lim \limits_{t \to \infty} P_i(t).$

Our assumptions allow to obtain the equations for the steady-state
probabilities by the standard manner [9]. For example the equations
corresponding to the states $E_1$ and $E_2$ are
$$-2(\rho_1\mu_1+\rho_2\mu_2) p_1+(1-\rho^2)\mu_1p_2+(1-\rho^2)\mu_2p_3=0,
\eqno(1)$$
$$-((1-\rho^2)\mu_1+2(\rho_1\mu_1+\rho_2\mu_2))p_2+
\mu_1 p_4+2(1-\rho)\mu_1 p_6+(1-\rho)\mu_2 p_7=0.\eqno(2)$$

The system for steady-states probabilities consists of the 45 equations,
corresponding to 45 system states, and the normalizing equation
$$\sum_{i=1}^{45}p_i=1.\eqno(3)$$

\subsection*{4. Macro-characteristics of flow}

According to the Markov processes theory each equations for the appropriate
state is a consequence of the other ones. The system of the 45 equations
and the normalizing equation has the only solution
${\overline p}_i,$ $i=1,\dots,4.$ This solution yields the
approximate values of the steady state probabilities of the four cells.

As it has been mentioned above the density $\rho_k$ of the flow of particles
of the $k$th type flow satisfies the relations $\rho_k=\frac{m_k}{2N}.$
$k=1,2.$ On the other hand
$$\rho_1=\frac{1}{2}(p_4+2p_9+p_{10}+p_{12}+p_{13}+p_{16}+p_{17}+p_{20}+
p_{21}+p_{22}+p_{23}+2p_{28}+$$
$$2p_{29}+p_{30}+p_{31}+p_{32}+p_{33}+2p_{36}+2p_{37}+2p_{38}+p_{39}+p_{40}+
p_{41}+p_{42}),\eqno(4)$$
$$\rho_2=\frac{1}{2}(p_5+p_{10}+2p_{11}+p_{14}+p_{15}+p_{18}+p_{19}+p_{24}+
p_{25}+p_{26}+p_{27}+p_{30}+$$
$$p_{31}+p_{32}+p_{33}+2p_{34}+2p_{35}+p_{39}+p_{40}+p_{41}+p_{42}+2p_{43}+
2p_{44}+2p_{45}).\eqno(5)$$

Let us prove the relations (4) and (5). The pair of cells
$(i,1),$ $(i,2)$ is identical with the pair $(j,1),$ $(j,2),$
$1\le i,j\le N.$ So the mean number of particles of $k$th type
$(k=1,2)$ in each pair equals $M_k/N,$ i.e. equals $2\rho_k$. On the other
hand the mean number of the particles of the $k$th type
in the pair of cells is equal to the sum of probabilities of the states of
the considerd four cells in that there is exactly one particle in the pair
of cells and the double probabilities of the states with two
particles in this pair of cells. Hence the equations (4) and (5) are true.

Let ${\overline \rho}_k$ be the value of $\rho_k$ which satisfies the
relations (4) and (5) if in this relations the approximate values
$\overline p_i$ substitute for exact values $p_i$.

Let mean number $q_k$ of particles of the $k$th type $(k=1,2)$ that leave
the pair of cells (1,1), (1,2) be called the particles of $k$th type
flow intensity and mean number $h_k$ of particles of the $k$th type leaving
the pair cells with a change of lane be called the changes intensity of the
particles of the $k$th type.

There is one particle, which can move, in the pair of cells (1,1) and (1,2),
when the state of the four cells is one from the states $E_4,$ $E_{10},$
$E_{12},$ $E_{13},$ $E_{16},$
$E_{17},$ $E_{28},$ $E_{29},$ $E_{30},$ $E_{31}.$ If the state of the four
cells is $E_9$ then there are two particles which can move. Hence the value
of $q_1$ can be calculated as
$$q_1=(p_4+2p_9+p_{10}+p_{12}+p_{13}+p_{16}+p_{17}+
p_{28}+p_{29}+p_{30}+p_{31})\mu_1. \eqno(6)$$

Similarly we obtain the relations
$$h_1=(p_{12}+p_{13})\mu_1,\eqno(7)$$
$$q_2=(p_5+p_{10}+2p_{11}+p_{14}+p_{15}+p_{18}+p_{19}+
p_{32}+p_{33}+p_{34}+p_{35})\mu_ 2,\eqno(8)$$
$$h_2=(p_{14}+p_{15})\mu_2.\eqno(9)$$

Let $\overline q_k,$ $\overline h_k$ $(k=1,2)$ be values of $q_k$ and $h_k$,
which are obtained if in equations (6)--(9) $p_i$ substitutes for
$\overline p_i$.

If $\mu_1>\mu_2$ then $\bar q_1$ is some greater than the value $q_1$
obtained by simulation and in contrast $\bar q_2$ is some smaller then
the value $q_2$ obtained by simulation. As it be shown below the difference
between the calculated approximate values and data obtained by simulation
is smaller if values $\tilde q_k,$ $\tilde h_k$ determined by means of the 
correcting coefficient substitute for values $q_k,$ $h_k$
$$\tilde q_k=(\rho_k/\bar \rho_k)\bar q_k,\ k=1,2,\eqno(10)$$
$$\tilde h_k=(\rho_k/\bar \rho_k)\bar h_k,\ k=1,2.\eqno(11)$$

Let us explain why the correcting coefficient improves he approximation.
Let the ratio of the mean number of transitions of particles of the $kth$
type to number of the particles of this type be called the velocity of
the particles of the $k$th
type. Let the velocity of particles of the $k$th type be $v_k.$ Then
$$v_k=q_k/(2\rho_k),\ k=1,2.\eqno(12)$$

Because of the relation $q_k=2\rho_k\cdot v_k$ the obtained approximate
value of the particles of the $k$th type differs from the exact value by
the coefficient ${\bar q}_2/\rho$. Coefficient $\rho/{\bar q}_2$ compensates
the error.

Suppose
$$\bar v_k=\bar q_k/(2\rho_k),\eqno(13)$$
$$\tilde v_k=\tilde q_k/(2\rho_k).\eqno(14)$$

\subsection*{5. Approximations by means of Bernoulli scheme}

Let us consider another manner of evaluation of the steady-state probabilities
and the macro-characteristics that is more simple than the manner described
above. Assume that the probability of the state of each cell is independent
of the states of other cells, and the cells with probability  $\rho_1$
contains a particle of the first type, with probability  $\rho_2$ contains
a particle of the second type and with probability $1-\rho$ the cell is
empty. Let $\hat p_i$ $(i=1,2,\dots 45),$ $\hat v_k,$ $\hat q_k,$
$\hat h_k$ $(k=1,2)$ be values of the appropriate characteristics that are
obtained when these assumptions are used.

Then $\hat p_1=(1-\rho)^4,$ $\hat p_2=\hat p_4=2\rho_1(1-\rho)^3,$
$\hat p_3=\hat p_5=2\rho_2(1-\rho)^3,$
$\hat p_6=\hat p_9=\rho_1^2(1-\rho)^2,$
$\hat p_7=\hat p_{10}=\hat p_{13}=\hat p_{14}=\hat p_{17}=\hat p_{18}=
2\rho_1\rho_2(1-\rho)^2,$
$\hat p_8=\hat p_{11}=\rho_2^2(1-\rho)^2,$
$\hat p_{12}=\hat p_{16}=2\rho_1^2(1-\rho)^2,$
$\hat p_{15}=\hat p_{19}=2\rho_2^2(1-\rho)^2,$
$\hat p_{20}=\hat p_{28}=2\rho_1^3(1-\rho),$
$\hat p_{21}=\hat p_{22}=\hat p_{24}=\hat p_{29}=\hat p_{30}=
\hat p_{32}=2\rho_1^2\rho_2(1-\rho),$
$\hat p_{23}=\hat p_{25}=\hat p_{26}=\hat p_{31}=\hat p_{33}=
\hat p_{34}=2\rho_1\rho_2^2(1-\rho),$
$\hat p_{27}=\hat p_{35}=2\rho_2^3(1-\rho),$ $\hat p_{36}=\rho_1^4,$
$\hat p_{37}=\hat p_{39}=2\rho_1^3\rho_2,$
$\hat p_{38}=\hat p_{43}=\rho_1^2\rho_2^2,$
$\hat p_{40}=\hat p_{41}=2\rho_1^2\rho_2^2,$
$\hat p_{42}=\hat p_{44}=2\rho_1\rho_2^3,$
$\hat p_{45}=\rho_2^4;$
$\hat v_k=(1-\rho)(1+\rho-\rho^2)\mu_k,\ k=1,2,$
$\hat q_k=2\rho_k(1-\rho)(1+\rho-\rho^2)\mu_k,\ k=1,2,$
$\hat h_k=2\rho_k\rho(1-\rho)^2\mu_k,$ $k=1,2.$

\subsection*{6. Sistem solution and results of calculation}

Suppose $\rho_1=\rho_2=0.25.$

The calculated approximate values are represented in table 1. These values
are compared to values obtained by simulation. The symbols for values
which have been obtained by simulations are marked by star. The duration of the simulation
interval is equal to 12000 time units, $N=500.$

\begin{table}[htbp]
\caption{State probabilities,$\rho_1=\rho_2=0.25$} 
\vskip 0.1cm
\footnotesize
\begin{tabular}{|c||c|c|c||c|c|c||c|c|c|}
\hline
&\multicolumn{3}{|c||}{$\mu_1=2,$ $\mu_2=1$}
& \multicolumn{3}{|c||}{$\mu_1=3,$ $\mu_2=1$}
&\multicolumn{3}{|c|}{$\mu_1=4,$ $\mu_2=1$} \\
\hline
$i$ &$\bar p_ i$ & $p^*_i$ & $\hat p_i$ & $\bar p_i$ & $p^*_i$ & $\hat p_i$ & $\bar p_i$ & $p^*_i$ & $\hat p_i$\\
\hline
1  &     0.091 &  0.098 &  0.063 &  0.094 &  0.005 &  0.063 &  0.097 &  0.133 &  0.063 \\
2  &     0.064 &  0.049 &  0.063 &  0.067 &  0.038 &  0.063 &  0.069 &  0.037 &  0.063 \\
3  &     0.053 &  0.072 &  0.063 &  0.050 &  0.069 &  0.063 &  0.047 &  0.082 &  0.063 \\
4  &     0.061 &  0.032 &  0.063 &  0.061 &  0.019 &  0.063 &  0.062 &  0.014 &  0.063 \\
5  &     0.064 &  0.087 &  0.063 &  0.066 &  0.090 &  0.063 &  0.067 &  0.099 &  0.063 \\
6  &     0.026 &  0.023 &  0.016 &  0.028 &  0.021 &  0.016 &  0.029 &  0.022 &  0.016 \\
7  &     0.041 &  0.040 &  0.031 &  0.040 &  0.037 &  0.031 &  0.039 &  0.035 &  0.031 \\
8  &     0.014 &  0.019 &  0.016 &  0.011 &  0.015 &  0.016 &  0.008 &  0.015 &  0.016 \\
9  &     0.022 &  0.011 &  0.016 &  0.021 &  0.006 &  0.016 &  0.021 &  0.004 &  0.016 \\
10 &     0.045 &  0.042 &  0.031 &  0.045 &  0.035 &  0.031 &  0.045 &  0.027 &  0.031 \\
11 &     0.023 &  0.032 &  0.016 &  0.023 &  0.039 &  0.016 &  0.023 &  0.038 &  0.016 \\
12 &     0.012 &  0.010 &  0.031 &  0.013 &  0.008 &  0.031 &  0.013 &  0.007 &  0.031 \\
13 &     0.015 &  0.014 &  0.031 &  0.018 &  0.013 &  0.031 &  0.019 &  0.012 &  0.031 \\
14 &     0.008 &  0.008 &  0.031 &  0.006 &  0.006 &  0.031 &  0.005 &  0.005 &  0.031 \\
15 &     0.011 &  0.012 &  0.031 &  0.011 &  0.011 &  0.031 &  0.010 &  0.012 &  0.031 \\
16 &     0.033 &  0.023 &  0.031 &  0.033 &  0.017 &  0.031 &  0.033 &  0.013 &  0.031 \\
17 &     0.025 &  0.023 &  0.031 &  0.021 &  0.015 &  0.031 &  0.019 &  0.012 &  0.031 \\
18 &     0.034 &  0.036 &  0.031 &  0.035 &  0.036 &  0.031 &  0.036 &  0.031 &  0.031 \\
19 &     0.025 &  0.037 &  0.031 &  0.021 &  0.037 &  0.031 &  0.018 &  0.037 &  0.031 \\
20 &     0.023 &  0.025 &  0.016 &  0.021 &  0.026 &  0.016 &  0.021 &  0.027 &  0.016 \\
21 &     0.014 &  0.015 &  0.016 &  0.014 &  0.015 &  0.016 &  0.013 &  0.015 &  0.016 \\
22 &     0.025 &  0.023 &  0.016 &  0.032 &  0.027 &  0.016 &  0.036 &  0.027 &  0.016 \\
23 &     0.018 &  0.017 &  0.016 &  0.018 &  0.018 &  0.016 &  0.017 &  0.017 &  0.016 \\
24 &     0.010 &  0.014 &  0.016 &  0.007 &  0.014 &  0.016 &  0.006 &  0.010 &  0.016 \\
25 &     0.007 &  0.010 &  0.016 &  0.005 &  0.008 &  0.016 &  0.005 &  0.005 &  0.016 \\
26 &     0.013 &  0.012 &  0.016 &  0.012 &  0.011 &  0.016 &  0.011 &  0.008 &  0.016 \\
27 &     0.009 &  0.011 &  0.016 &  0.006 &  0.007 &  0.016 &  0.005 &  0.007 &  0.016 \\
28 &     0.016 &  0.014 &  0.016 &  0.016 &  0.016 &  0.016 &  0.016 &  0.015 &  0.016 \\
29 &     0.023 &  0.018 &  0.016 &  0.027 &  0.018 &  0.016 &  0.030 &  0.015 &  0.016 \\
30 &     0.009 &  0.009 &  0.016 &  0.006 &  0.008 &  0.016 &  0.005 &  0.005 &  0.016 \\
31 &     0.012 &  0.009 &  0.016 &  0.011 &  0.008 &  0.016 &  0.010 &  0.008 &  0.016 \\
32 &     0.017 &  0.021 &  0.016 &  0.018 &  0.026 &  0.016 &  0.018 &  0.025 &  0.016 \\
33 &     0.024 &  0.023 &  0.016 &  0.029 &  0.033 &  0.016 &  0.032 &  0.036 &  0.016 \\
34 &     0.009 &  0.008 &  0.016 &  0.006 &  0.010 &  0.016 &  0.005 &  0.006 &  0.016 \\
35 &     0.013 &  0.015 &  0.016 &  0.011 &  0.013 &  0.016 &  0.010 &  0.012 &  0.016 \\
36 &     0.007 &  0.009 &  0.004 &  0.008 &  0.019 &  0.004 &  0.008 &  0.027 &  0.004 \\
37 &     0.020 &  0.017 &  0.008 &  0.026 &  0.028 &  0.008 &  0.030 &  0.032 &  0.008 \\
38 &     0.013 &  0.008 &  0.004 &  0.020 &  0.014 &  0.004 &  0.025 &  0.015 &  0.004 \\
39 &     0.007 &  0.014 &  0.008 &  0.005 &  0.017 &  0.008 &  0.004 &  0.016 &  0.008 \\
40 &     0.010 &  0.009 &  0.008 &  0.009 &  0.013 &  0.008 &  0.009 &  0.012 &  0.008 \\
41 &     0.010 &  0.012 &  0.008 &  0.009 &  0.012 &  0.008 &  0.007 &  0.013 &  0.008 \\
42 &     0.014 &  0.008 &  0.008 &  0.014 &  0.011 &  0.008 &  0.014 &  0.008 &  0.008 \\
43 &     0.002 &  0.003 &  0.004 &  0.001 &  0.004 &  0.004 &  0.001 &  0.003 &  0.004 \\
44 &     0.005 &  0.005 &  0.008 &  0.003 &  0.004 &  0.008 &  0.002 &  0.002 &  0.008 \\
45 &     0.003 &  0.003 &  0.004 &  0.002 &  0.002 &  0.004 &  0.002 &  0.002 &  0.004 \\
\hline
\end{tabular}
\end{table}

\large

We do not present the tables of calculations for $\rho_1=0.54,$
$\rho_2=0.06$ and $\rho_1=0.08,$ $\rho_2=0.72.$

The dependence of state probabilities on state numbers is represented on
fig 2-10.

\begin{figure}[h]
\centerline{\includegraphics[width=14cm]{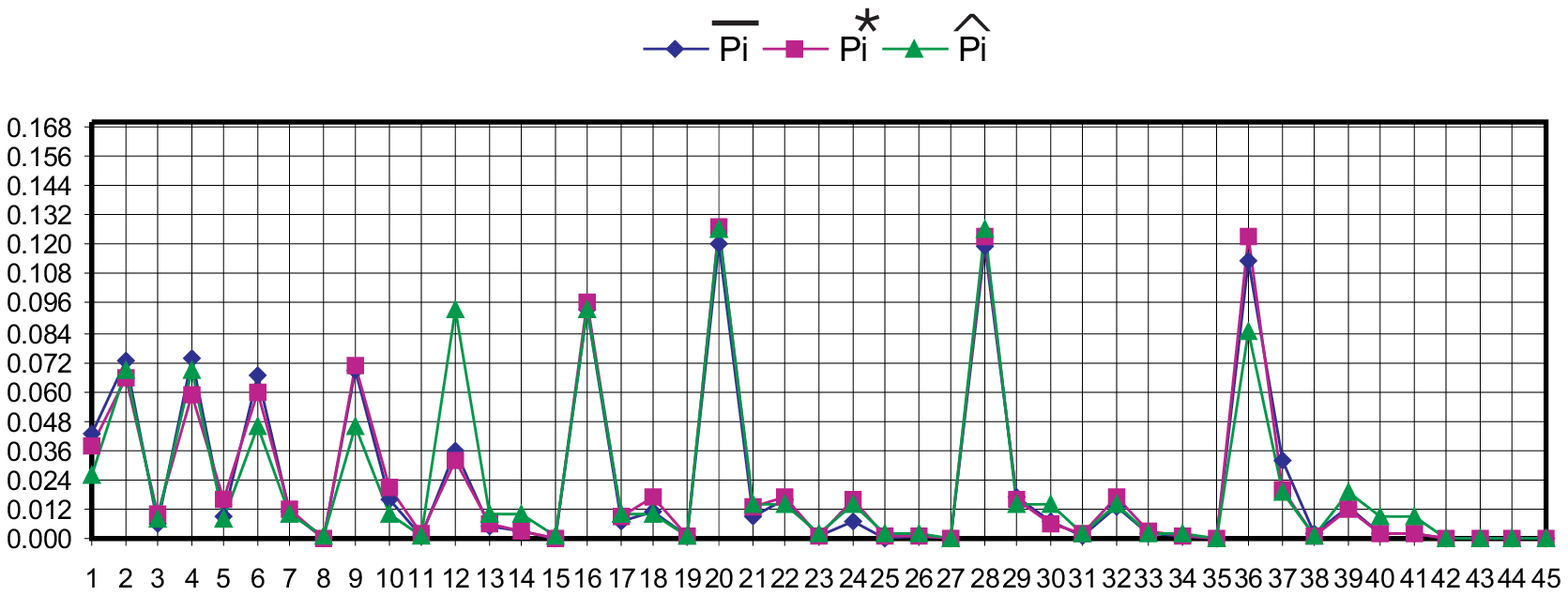}}
\caption{State probabilities for $\rho_1=\rho_2=0.25,$ $\mu_1=2,$ $\mu_2=1$
}
\label{mm21}
\end{figure}

\begin{figure}[h]
\centerline{\includegraphics[width=14cm]{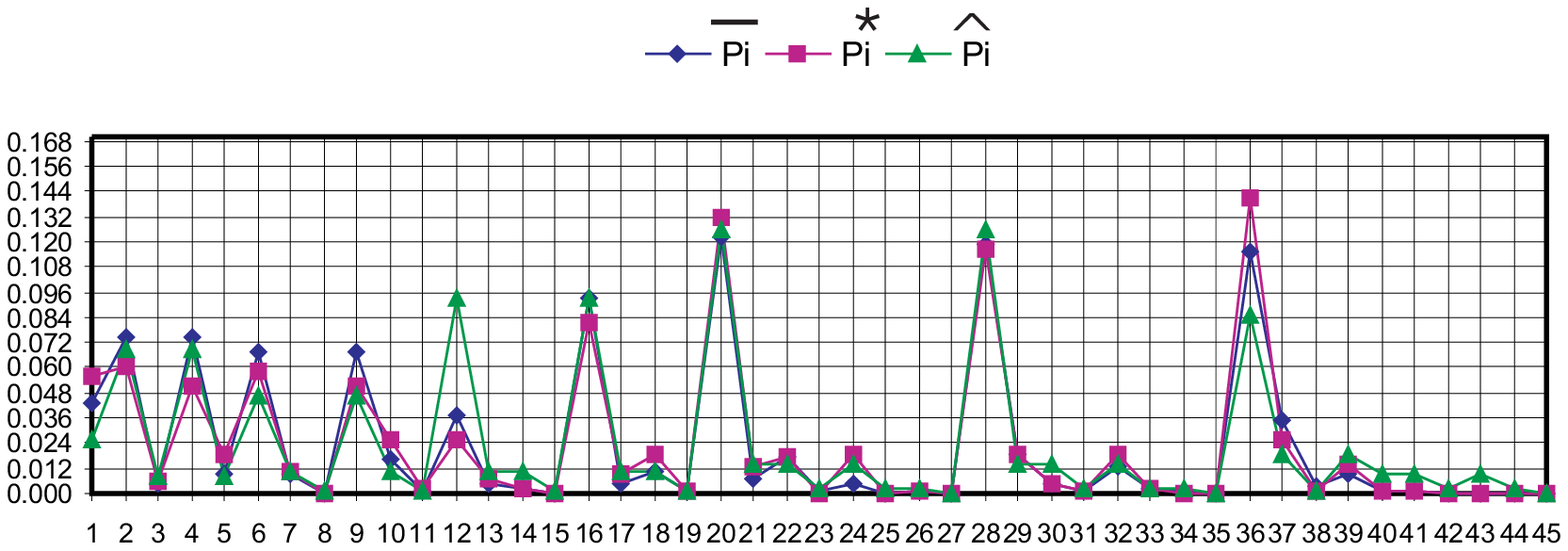}}
\caption{State probabilities for $\rho_1=\rho_2=0.25,$ $\mu_1=3,$ $\mu_2=1$
}
\label{mm31}
\end{figure}

\begin{figure}[h]
\centerline{\includegraphics[width=14cm]{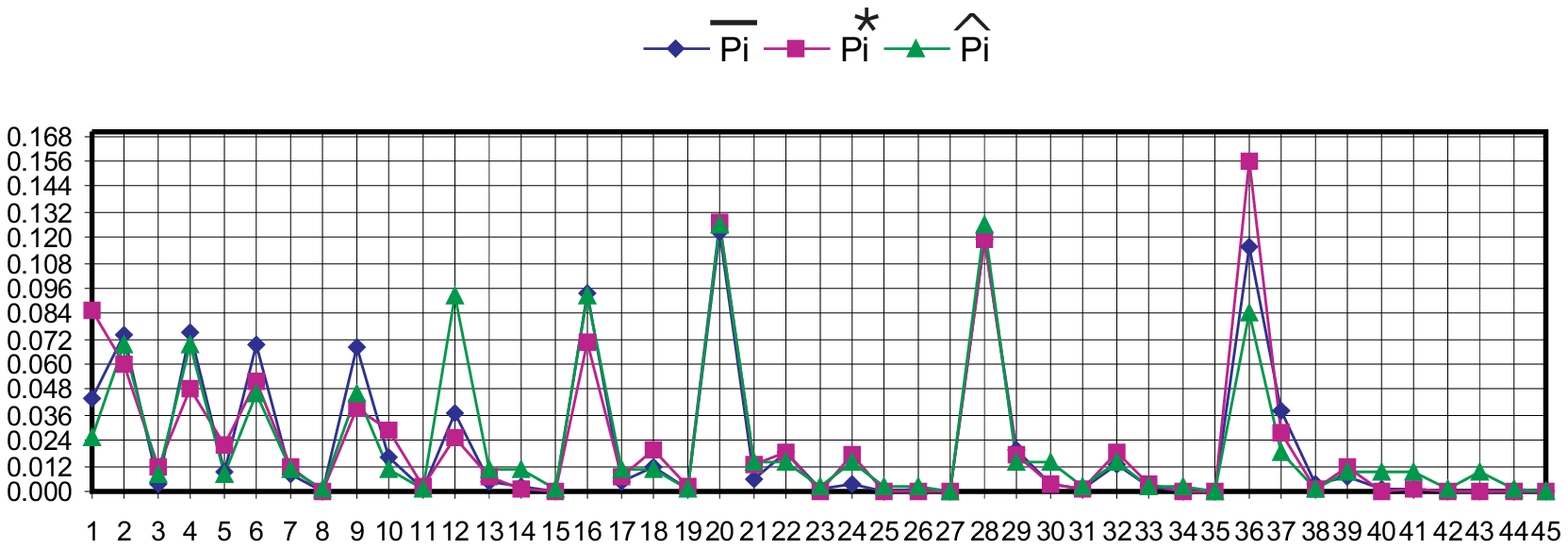}}
\caption{State probabilities for $\rho_1=\rho_2=0.25,$ $\mu_1=4,$ $\mu_2=1$
}
\label{mm41}
\end{figure}



\begin{figure}[h]
\centerline{\includegraphics[width=14cm]{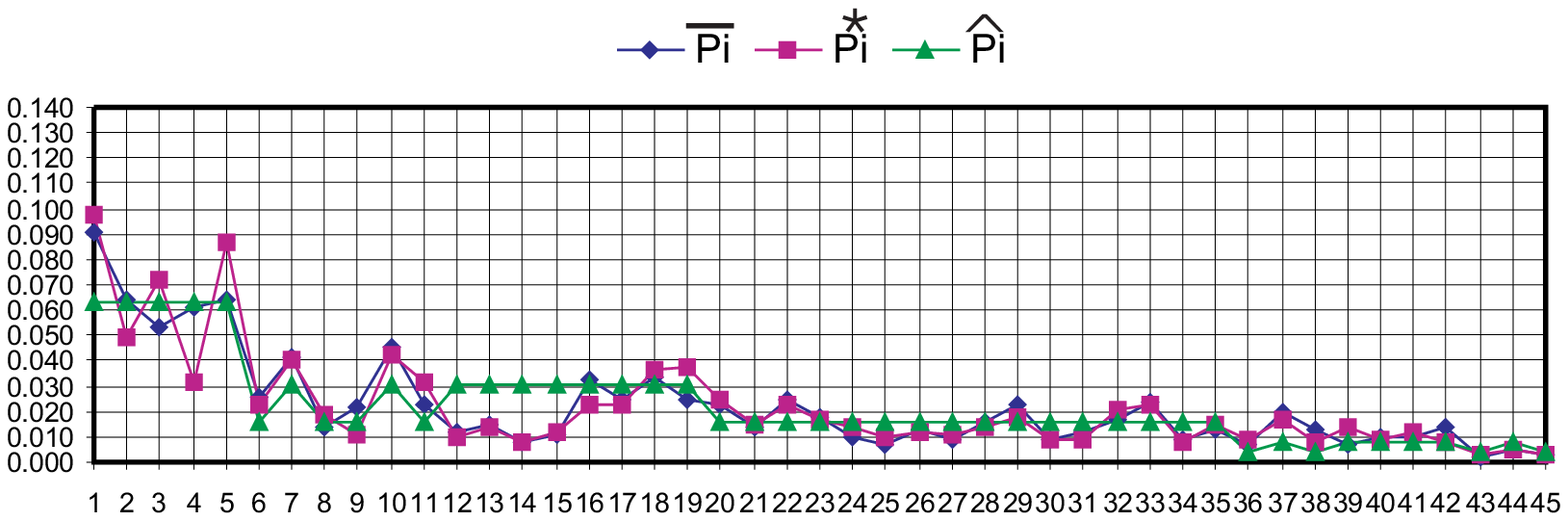}}
\caption{State probabilities for $\rho_1=0.54,$ $\rho_2=0.06,$ $\mu_1=2,$ $\mu_2=1$
}
\label{mm41}
\end{figure}


\begin{figure}[h]
\centerline{\includegraphics[width=14cm]{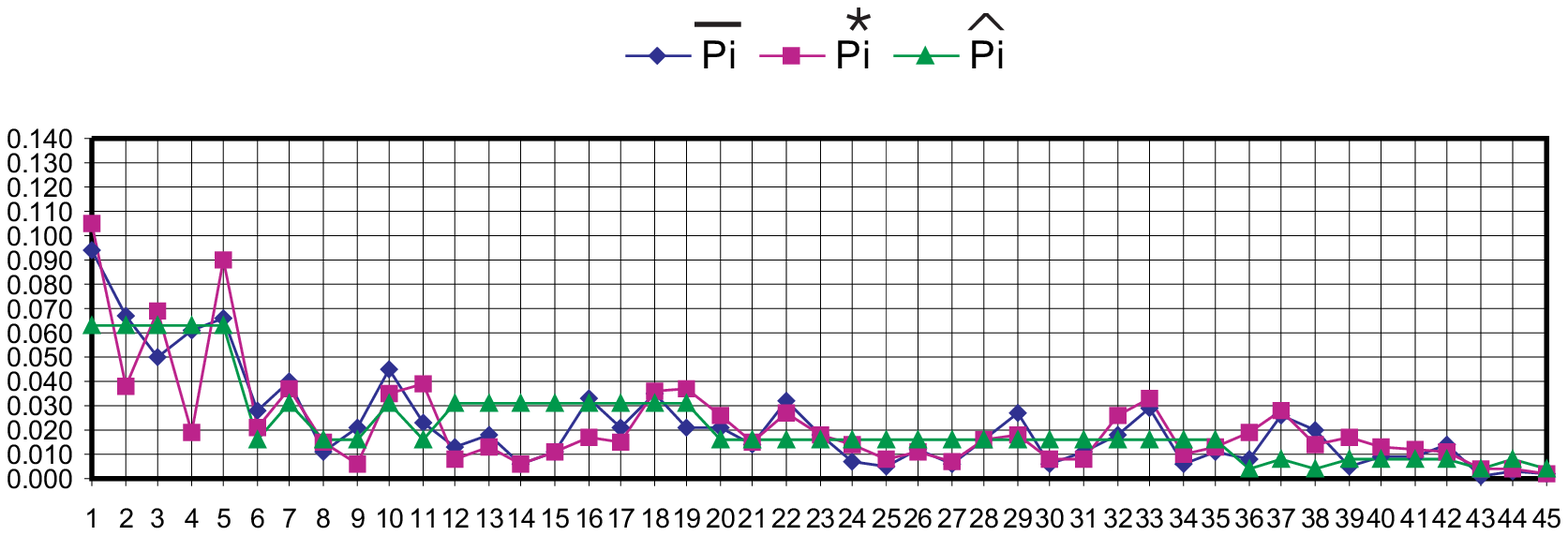}}
\caption{State probabilities for $\rho_1=0.54,$ $\rho_2=0.06,$ $\mu_1=3,$ $\mu_2=1$
}
\label{b31}
\end{figure}

\begin{figure}[h]
\centerline{\includegraphics[width=14cm]{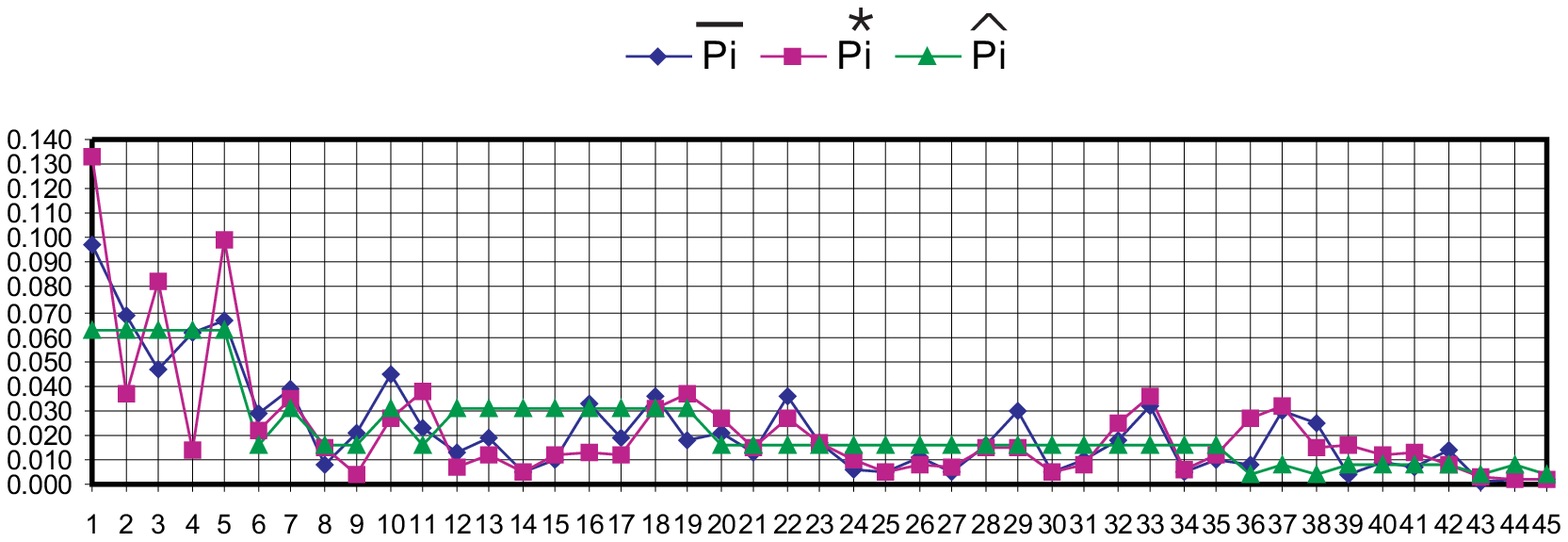}}
\caption{State probabilities for $\rho_1=0.54,$ $\rho_2=0.06,$ $\mu_1=4,$ $\mu_2=1$
}
\label{b41}
\end{figure}

\begin{figure}[h]
\centerline{\includegraphics[width=14cm]{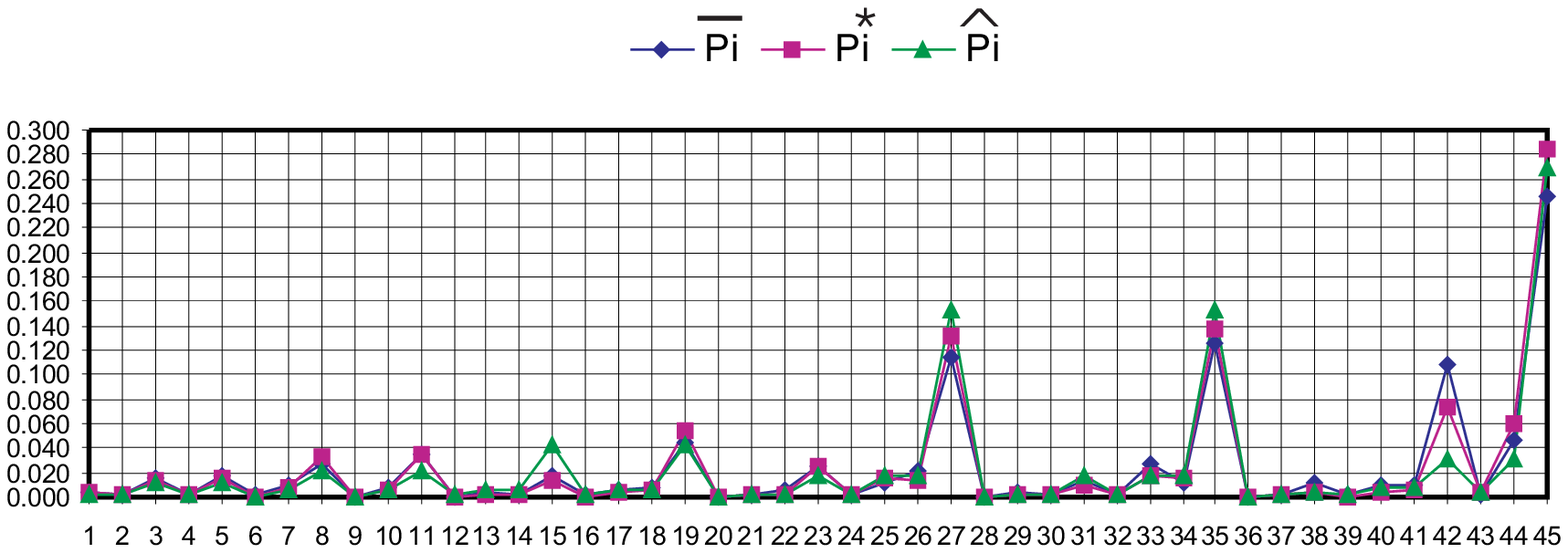}}
\caption{State probabilities for $\rho_1=0.08,$ $\rho_2=0.72,$ $\mu_1=2,$ $\mu_2=1$
}
\label{b21}
\end{figure}

\begin{figure}[h]
\centerline{\includegraphics[width=14cm]{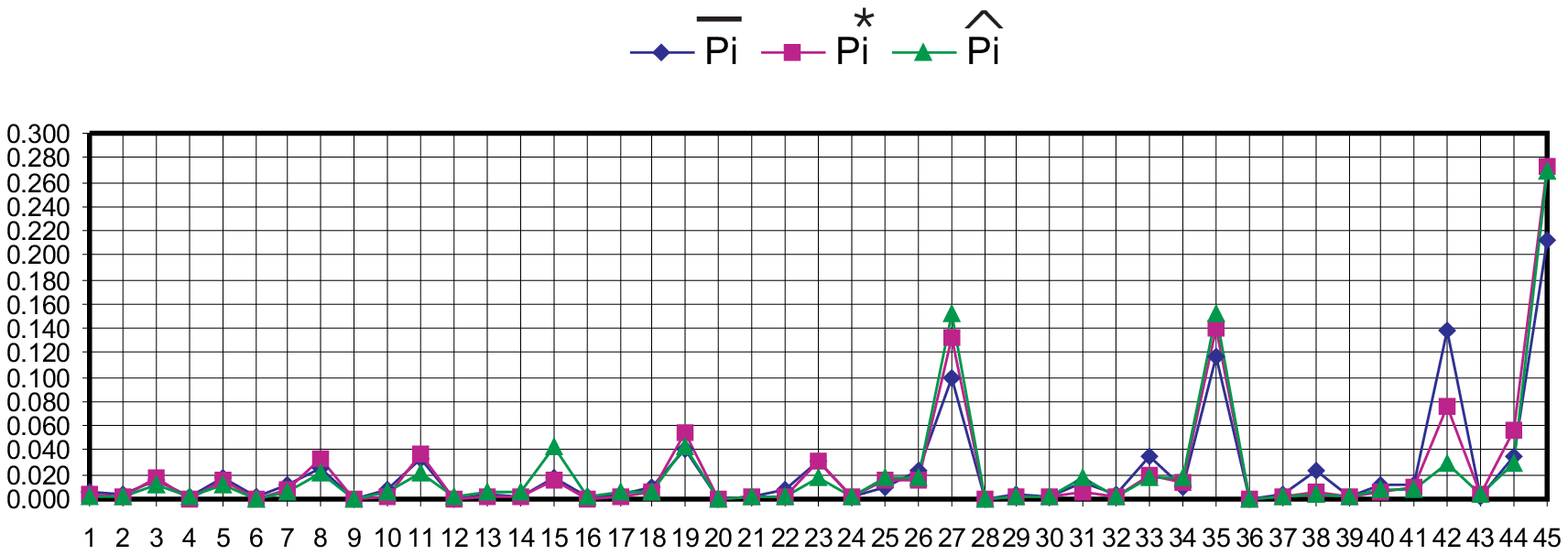}}
\caption{State probabilities for $\rho_1=0.08,$ $\rho_2=0.72,$ $\mu_1=3,$ $\mu_2=1$
}
\label{b21}
\end{figure}

\begin{figure}[h]
\centerline{\includegraphics[width=14cm]{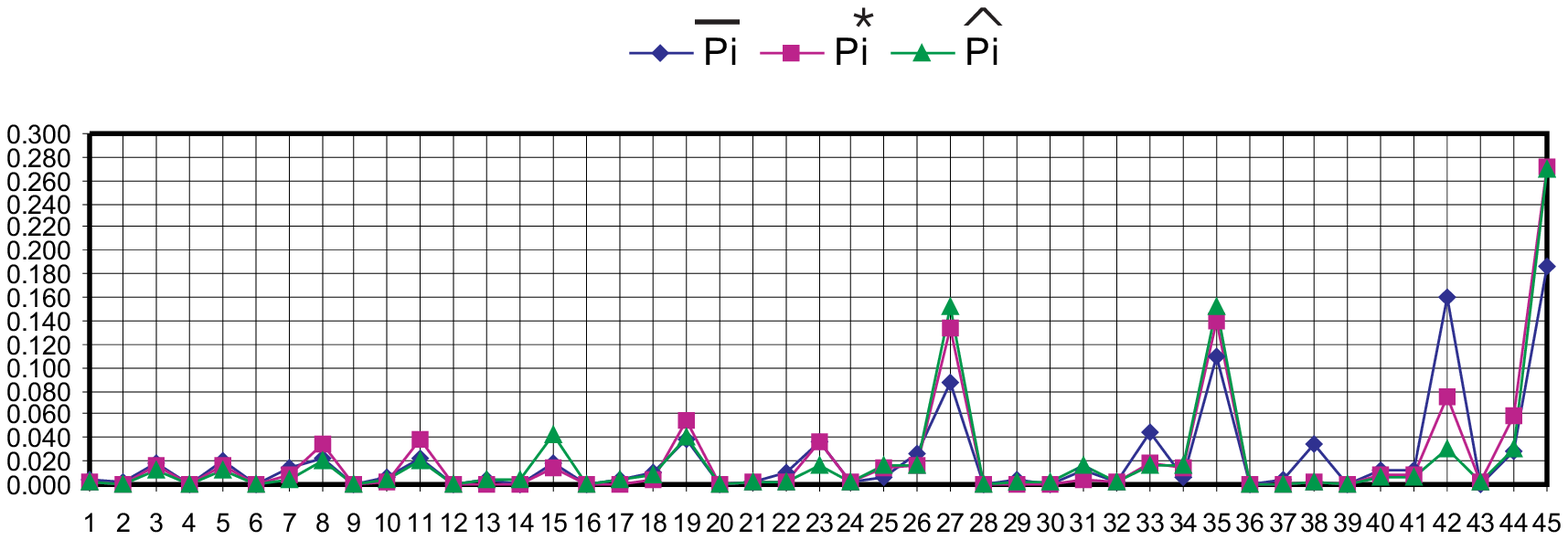}}
\caption{State probabilities for $\rho_1=0.08,$ $\rho_2=0.72,$ $\mu_1=4,$ $\mu_2=1$
}
\label{c41}
\end{figure}

The calculated values of macro-characteristics are represented in table 2.

\begin{table}[htbp]
\caption{Macro-characteristics (the error that is relative to values obtained by
simulation} 
\vskip 0.1cm
\small
\hspace{-1cm}%
\begin{tabular}{|c|c|c|c|c|c|c|}
\hline
\multicolumn{1}{|c|}{}
&\multicolumn{2}{|c|}{$\mu_1=2,$ $\mu_2=1$}
& \multicolumn{2}{|c|}{$\mu_1=3,$ $\mu_2=1$}
&\multicolumn{2}{|c|}{$\mu_1=4,$ $\mu_2=1$} \\
\hline
$k$           &$1$    & $2$       & $1$   & $2$   & $1$   & $2$   \\
\hline
\multicolumn{7}{|c|}{$\rho_1=\rho_2=0.25$}\\
\hline
$q^*_k$       & 0.432         &0.353        & 0.507       &0.375     & 0.544          &0.366      \\
$\tilde q_k$  & 0.512 (+19) &0.337$(-5)$  & 0.717(+42)&0.361 $(-4)$& 0.923 (+70) &0.372(+2)       \\
$\hat q_k$    & 0.625 (+45) &0.313$(-12)$ & 0.938(+85)&0.313 $(-17)$& 1.250(+130)&0.313$(-15)$      \\
$\bar q_k$    & 0.590 (+37) &0.296$(-16)$ & 0.879(+74)&0.294 $(-22)$& 1.176(+116)&0.292$(-20)$        \\\hline
$h^*_k$       & 0.048         &0.020        & 0.063       &0.017      & 0.076          &0.017      \\
$\tilde h_k$  & 0.047 $(-2)$  &0.022 (+8) & 0.076(+20)&0.021 (+23)& 0.101$(+32)$ &0.019$(+12)$ \\
$\hat h_k$    & 0.125 (+160) &0.063(+212)&0.188(+198)&0.063(+268)& 0.250(+229)&0.063(+268)      \\
$\bar h_k$    & 0.054 (+13) &0.019 $(-5)$  &0.093(+47)&0.017 (0)   & 0.128 (+69)&0.015$(-12)$        \\\hline
$v^*_k$       & 0.864         &0.706        & 1.014       &0.750      & 1.088          &0.732      \\
$\tilde v_k$  & 1.024 (+19) &0.674 $(-5)$ & 1.434 (+42)&0.722$(-4)$  & 1.846 (+70)&0.743 (+2)     \\
$\hat v_k$    & 1.250 (+45) &0.625 $(-12)$& 1.875 (+85)&0.625$(-17)$ & 2.500 (+130) &0.625 $(-15)$    \\
$\bar v_k$    & 1.180 (+37) &0.592 $(-16)$& 1.758 (+74)&0.588$(-22)$ & 2.352 (+116)&0.584$(-20)$       \\
\hline
\multicolumn{7}{|c|}{$\rho_1=0.54,$ $\rho_2=0.06$}\\
\hline
$q^*_k$       &1.024        &0.083       & 1.335       &0.092      &1.624      &0.101       \\
$\tilde q_k$  &0.989 (-3) &0.076$(-8)$ & 1.465 (+10) &0.084$(-8)$  &1.944(+20) &0.090$(-11)$       \\
$\hat q_k$    &1.071 (+4) &0.060$(-28)$ & 1.607(+20) &0.060$(-35)$ &2.143(+32) &0.060$(-41)$       \\
$\bar q_k$    &1.026(+19) &0.058$(-30)$ & 1.530(+15)&0.057$(-38)$&2.048(+26)&0.056$(-45)$       \\\hline
$h^*_k$       &0.076        &0.003       & 0.099       &0.002      &0.128  &0.001       \\
$\tilde h_k$  &0.079(+4)  &0.004(+30) & 0.121(+22) &0.003 (+48) &0.159(+24)  &0.003(+220)        \\
$\hat h_k$    &0.207(+172)&0.012(+284)& 0.311(+214)&0.012 (+476)&0.415(+224) &0.012(+1052)       \\
$\bar h_k$    &0.082(+8)&0.003  (0)   & 0.126(+27) &0.002 (0)  &0.168(+31)  &0.002(+100)        \\\hline
$v^*_k$       &0.948        &0.692       & 1.236       &0.767      &1.504      &0.842       \\
$\tilde v_k$  &0.915$(-3)$  &0.630$(-8)$ & 1.356 (+10)&0.704$(-8)$       &1.800(+20) &0.747$(-11)$       \\
$\hat v_k$    &0.992(+4)  &0.496$(-28)$ & 1.488(+20)&0.496$(-35)$      &1.984(+32) &0.496$(-41)$       \\
$\bar v_k$    &0.950(+19) &0.483$(-30)$ & 1.417(+15)&0.475$(-38)$      &1.896(+26)&0.467$(-45)$       \\
\hline
\multicolumn{7}{|c|}{$\rho_1=0.08,$ $\rho_2=0.72$}\\
\hline
$q^*_k$       &0.044     &0.334          & 0.039       &0.344        &0.048         &0.346     \\
$\tilde q_k$  &0.045(+2)  &0.351(+5) & 0.050(+28) &0.371(+8) &0.056(+17)  &0.374(+8)  \\
$\hat q_k$    &0.074(+69) &0.334(0)  & 0.111(+186)&0.334$(-3)$ &0.149(+210) &0.334$(-3)$  \\
$\bar q_k$    &0.072(+64) &0.328$(-2)$ & 0.105(+169)&0.325$(-5)$ &0.140(+192)  &0.305$(-12)$ \\\hline
$h^*_k$       &0.002        &0.015       & 0.003       &0.017         &0.004        &0.016      \\
$\tilde h_k$  &0.004(+87) &0.020(+36) & 0.004(+43)  &0.023(+34) &0.006(+60)  &0.025(+54)       \\
$\hat h_k$    &0.010(+412)&0.046(+207)& 0.015(+412) &0.046(+171)&0.021(+412) &0.046(+188)      \\
$\bar h_k$    &0.006(+200)&0.019(+27) & 0.009(+200) &0.020(+18) &0.016(+300) &0.020(+25)       \\\hline
$v^*_k$       &0.275     &0.232          & 0.244       &0.239         &0.300        &0.240      \\
$\tilde v_k$  &0.280 (+2) &0.244 (+5)& 0.313(+28) &0.258(+8)  &0.350(+17) &0.260(+8)  \\
$\hat v_k$    &0.464 (+69)&0.232 (0) & 0.696(+186)&0.232$(-3)$  &0.928(+210)&0.232$(-3)$ \\
$\bar v_k$    &0.450 (+64)&0.228 $(-2)$& 0.656(+169)&0.226$(-5)$  &0.875(+192) &0.212$(-12)$\\
\hline
\end{tabular}
\end{table}
\large

\subsection*{7. Analysis of approximation accuracy}

Let us compare the values
$\tilde q_k,$ $\tilde v_k,$ $\tilde h_k,$
$\bar q_k,$ $\bar v_k,$ $\bar h_k,$
$\hat q_k,$ $\hat v_k,$ $\hat h_k$
with the values obtained by simulation.

Suppose $\rho_1=\rho_2=0.25.$ The values ${\tilde q}_1$ $({\tilde v}_1)$
exceeds the values $q^*_1$ $(v^*_1).$ The relative error makes up the
values from 19\% to 70\% on the three considered set of data.
The value ${\bar q}_1$ $\bar (v_1)$ exceeds  the value $q^*_1$ $(v^*_1).$
The difference ${\bar q}_1$ and $q^*_1$ makes up from 37\% to 116\% of
$q^*_1.$ The value ${\hat q}_1$ $({\hat v}_1)$  exceeds the value
$q^*_1$ $(v^*_1).$ The relative error make up from 45\% to 113\%.

In the other cases the approximate values calculated according to formulas
(10), (11), (14) are also nearer to value obtained by simulation than the
values calculated by the two others manners, or in some cases the
formulas (10), (11), (14) yields approximation only few worse.

In each the three cases the value of  ${\tilde q}_2$
$({\tilde v}_2)$ difference the values  $q^*_2$ $(v^*_2)$ is not
greater than 5\% of $q^*_2.$  The value  ${\tilde h}_1$  difference
from the value $h^*_1$ not greater than 32\% of $h^*_1$. The value
$h^*_2$ exeeds $h^*_2.$ The relative error is not greater than 23\%.

Usually the relative error is the better the smaller is
the ratio of  $\mu_1$ to $\mu_2$ ($\mu_1>\mu_2$).

The values $\tilde p_1,$ $\tilde p_2,\dots,\tilde p_{45}$
that satisfies the system of linear equations
usually differs from the values obtained
simulation less than values calculated by means of Bernoulli scheme.

The analysis of calculation for the cases $\rho_1=0.54,$
$\rho_2=0.06,$ and $\rho_2=0.06,$ $\rho_1=0.72$, also
shows that meaning $\tilde v_k$,
$\tilde q_k$, $\tilde h_k$, are usually more close differs
from the values obtained by simulation less than values
$\bar v_k,$ $\bar q_k,$ $\bar h_k,$ $\hat v_k,$ $\hat q_k,$ $\hat h_k.$

For $\rho_1=0.54,$ $\rho_2=0.06,$ the largest difference between
${\tilde q}_k$ $({\tilde v}_k)$ and $q^*_k$ $(v^*_k)$
corresponds to the case $k=1,$ $\mu_1=4,$ $\mu_2=1$. In
this case the value  ${\tilde q}_1$ exceeds the value  $q^*_2.$
The relative error is not greater than  20\% .

For $\mu_1=0.08,$ $\mu_2=0.72$ the largest difference between
${\tilde q}_k$ $({\tilde v}_k)$ $q^*_k$ $(v^*_k)$
is reached in the case $k=1,$ $\mu_1=3,$ $\mu_2=1$. In this case the value
${\tilde q}_1$ $(\tilde v_1^*)$ exceeds values $q^*_1$
$(v^*_1).$ The relative error is not greater than 28\%.

\subsection*{8. Analysis of results of calculations of\\
macro-characteristics}

If  Bernoulli scheme approximation is used then calculated values of
macro-characteristics are proportional to the particle transitions intensity.

On the other hand it is clear intuitively
that the fast particles occur more often than slow ones in
the states in that their transitions are prevented by other
particles. Hence the ratio of velocity $(v_k)$ of
particle to its transition intensity $(\mu_k)$ for the fast particles
is less than for the slow particles. It is confirmed both by
simulation and calculations. The difference between
$v_2/\mu_2$ and $v_1/\mu_1$ is the greater the greater the $\mu _1/\mu_2$ is.
Because of that for fixed $\mu_2$  both value $v^*$ and value
$\tilde v_1$ calculated according to formulas (10), (14) increases slower
than linearly if $\mu_1$ increases.
But it can be noticed that for the valued obtained by
simulation the slowing down of increasing of the velocity
of the particle of the first type with increasing of
$\mu_1$ is more obvious. It can be noticed also if
$\mu_2$ is fixed and $\mu_1$ increases
then the values $\bar v_2$ and $v^*_2$ increases slowly and the value
${\bar v}_2$ is constant. For fixed $\mu_2$ and increasing $\mu_1$ the values
${\bar h}_1$ and $h^*_1$ increase approximately linearly.
The ratio of ${\bar h}_k/{\bar q}_k$ to
$h^*_ k/q^*_ k $ for  the fast particles is greater than for slow
ones (moving forward the fast particles change lane more often than slow
particles).

\subsection*{9. Conclusion}

It is presented a model of the traffic flow on two lanes. An
approximate manner is elaborated for calculation
the steady state probabilities and macro-characteristics of model.
The results of calculations by means of the elaborated method are
compared with results of simulation. This comparison
shows that the presented method yield rather good
approximation and describes the qualitative behavior
of the considered  characteristics.
\newpage
\vskip 0.5cm
\section*{ References}


1. Belyaev, Y.K. (1969)
'On Simple Traffic Model without Passing',
{\it Izv.  AN SSSR. S. Cybernetics (in Russian)}, N 3. p. 17-21.

2. Zele, U. (1972)
'Generation of Traffic Model without Passing',
{\it Izv. AN SSSR. S. Cybernetics (in Russian)} N 5. p. 100-103.

3. Schreckenberg M., Schadschneider A., Nagel K., Ito N. Discrete
stochastic models for traffic flow.~--- Phis. Rev., E~--- 1995.~---V. 51.~---
P. 2939~--- 2949.

4. Lukanin, V.N.,  Buslaev, A.P., Trofimenko, Yu.V., Yashina,
M.V. (1998) {\it Traffic Flows and Environment,}
{\it Moscow, INFRA-M (in Russian)}, p. 408.

5. Lukanin, V.N.,  Buslaev, A.P., Yashina, M.V. (2001)
{\it Traffic  Flows  and Environment - 2,}
{\it Moscow, INFRA-M (in Russian)}, p. 644.

6. Daganzo C.F. A finite difference approximation for the kinetic wave
model. Institute of Transportation Studies.~--- {\it Trans. Res.~---  1993.~---
V.~29B (4).}~--- P. 261~--- 276.

7. Helbing D. Modelling multilane traffic flow with queueing effects.~---
{\it Physics, A~--- 1997.~--- V. 242.}~--- P. 175~---~194.

8.  Belyaev Yu.K., Buslaev A.P., Seleznev~O.V., Tatashev~A.G.,
Yashina~M.V. (2002) Markov approximation of two-lane traffic model//
{\it Moscow, MADI-STU (in Russian).}~--- 32 p.

9. Karlin S. (1968) A first course in stochastic process.
{\it Academic press. New York
and London.}
\end{document}